\documentclass[
twocolumn,
bibnotes,
amsmath,amssymb,
aps,
prb,
floatfix,
longbibliography,
eprint,
]{revtex4-2}

\pdfoutput=1

\usepackage{xcolor,mathrsfs,dsfont}
\definecolor{darkblue}{RGB}{0,0,150}
\definecolor{nightblue}{RGB}{0,0,100}

\usepackage{graphicx,mathtools,bm}
\usepackage[
colorlinks,
citecolor=darkblue,
linkcolor=darkblue,
urlcolor=nightblue]{hyperref}

\usepackage[english]{babel}
\usepackage[babel,kerning=true,spacing=true]{microtype}

\renewcommand{\Re}{\ensuremath{\mathrm{Re}\,}}

\newcommand{\beq}{\begin{equation}}
\newcommand{\eeq}{\end{equation}}

\bibpunct{[}{]}{,}{n}{}{}
\makeatletter
\def\NAT@def@citea{\def\@citea{\NAT@separator}}
\makeatother

\begin{document}

\title{
Electrons flow like falling cats:\protect\\
Deformations and emergent gravity in quantum transport
}
\author{Tobias Holder}
\affiliation{Department of Condensed Matter Physics, Weizmann Institute of Science, Rehovot, Israel 76100
}

\date{\today}

\begin{abstract}
The effective low-energy excitations in a metallic or semimetallic crystalline system (i.e. electronic quasiparticles) always have a finite spatial extent. It is self-evident but virtually unexplored how the associated internal degrees of freedom manifest themselves in the quasiparticle response. 
Here, we investigate this question by illuminating an intimate connection between the theory of nonlinear response and the equations of motion of classical deformable bodies. 
This connection establishes that nth-order response in an external perturbation corresponds to nth-order derivatives of the quasiparticle motion, where the resulting motion is anomalous at every order due to the internal degrees of freedom. 
This new point of view predicts that 
quasiparticles necessarily move in an emergent curved spacetime, even in a homogeneous and defectless lattice.
We underscore these concepts using recent results on the second order electrical conductivity, elucidating the associated anomalous acceleration that the quasiparticle exhibits. 
Based on our observations, we predict the existence of an infinite series of anomalous components of the quasiparticle motion, and propose a new mapping between response theory in flat space and a gravitational theory for the center-of-mass coordinate. 
\end{abstract}

\maketitle

\section{Introduction}
There is no doubt neither in theory~\cite{Evers2008,Nayak2008,Marzari2012} nor from experiment~\cite{Avraham2018} that the low-energy electronic excitations in gapless condensed matter systems have a finite size.
Nevertheless, surprisingly little is know about the secular motion which arises when a moving wavepacket is squeezed or stretched as it traverses the lattice.
This is understandable, because at first sight, if a transient quantum mechanical state is spatially extended, this is not a sufficient criterion for the emergence of internal degrees of freedom because the spatially extended state could still mostly move like a rigid body. 
The question under which circumstances this is not the case and the wavepacket does change shape is rather subtle because the latter is necessarily a \emph{dynamical} phenomenon.
Despite the apparent sparsity of examples where a spatial deformation of the electronic quasiparticle seems to matter, here we propose that semiclassical quasiparticle motion must \emph{always} be described as the motion of a dynamically deforming body.

Two powerful arguments in favor of this proposition will be discussed here: Firstly, there are striking formal similarities between the expressions derived from the gauge theory of deformable bodies~\cite{Littlejohn1997} and the results obtained from canonical perturbation approaches~\cite{Parker2019,Gao2019,Holder2020}. Secondly, we comment on recent progress regarding the second order conductivity, where it is possible to make concrete and experimentally testable predictions based on our proposition.
\begin{figure}
    \centering
    \includegraphics[width=.9\columnwidth]{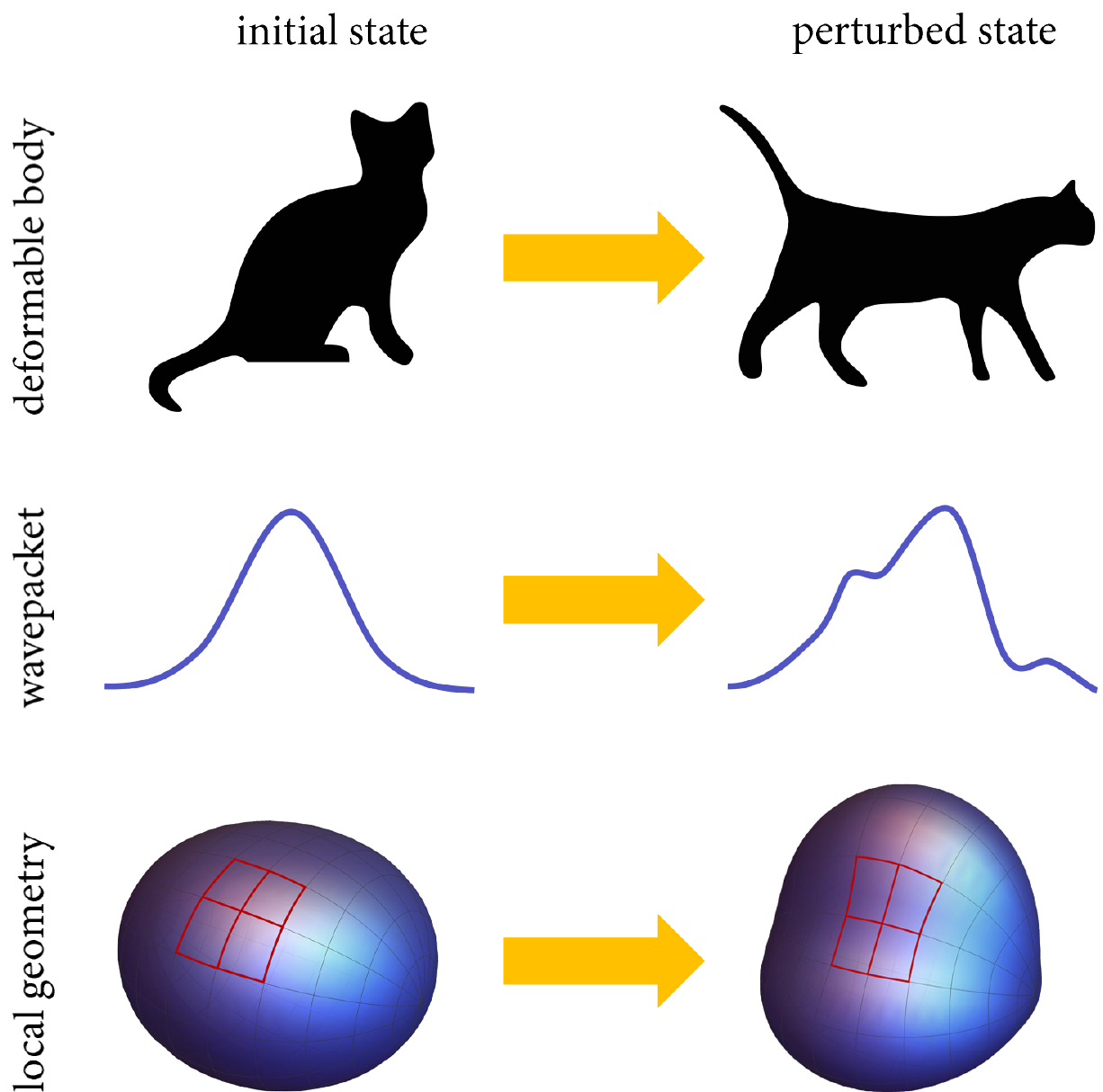}
    \caption{
    A moving quasiparticle (i.e. a wavepacket) necessarily suffers spatial deformations from the lattice potential. Therefore, the classical motion of deformable bodies is formally very similar to the motion of electrons in a periodic lattice.
    Because the wavepacket is dynamically deformed, the center-of-mass coordinate effectively moves in an emergent curved spacetime.}
    \label{fig:fig1}
\end{figure}
On top of that, one important example which supports our central proposition is actually well-known, which is the \emph{anomalous velocity}:
It has been appreciated in the last 20 years that the semiclassical equation of motion has to be slightly corrected if the system in consideration has a finite Berry curvature~\cite{Nagaosa2010,Xiao2010}. For a long time, this effect was considered somewhat odd (anomalous), given that the semiclassical approach examines the quasiparticle motion in terms of the center-of-mass coordinate of the wavepacket~\cite{Karplus1954}, and the need to accommodate this additional velocity component was considered one among many peculiarities of quantum transport~\cite{Nagaosa2010}.
Many valid interpretations to the appearance of the anomalous velocity have been given since~\cite{Nagaosa2010,Xiao2010,Dalibard2011,Manchon2015}, which we do not refute. 
In this work, we introduce a novel perspective, namely that the need to define a gauge field for the quasiparticle motion \emph{necessarily} means that the quasiparticle motion is (1) the motion of a deformable body, and that (2) the resulting semiclassical equations of motion are embedded in an emergent curved spacetime. 
We claim this applies in the bulk of a homogeneous material, and even at zero temperature and in the limit of vanishing disorder. 
In other words, the emergence of a curved spacetime is an intrinsic property of any lattice system that the low-energy excitations can be probed for.

Quasiparticle motion essentially resembles the tumbling motion of a falling quantum cat (cf.~Fig.~\ref{fig:fig1}):
When a deformable body like for example a cat moves, it not only changes position and rotation, but also its shape. The same holds for electronic excitations (wavepackets) which move in a periodic lattice. Because the spatial shape of the excited state changes as a function of time, the local geometry of the low-energy degrees of freedom of the excitations change, and this change in the local geometry gives rise to a nontrivial metric.

In the following, we first summarize our main results. Then we introduce the gauge theory of deformable bodies, using this language to expose the importance of the internal dynamics for the quasiparticle response.
We then give several examples for the anomalous acceleration and the resulting quasiparticle dynamics. We conclude with some conjectures about other promising applications of the framework presented here.

\section{Summary of results}

\begin{figure}
    \centering
    \includegraphics[width=\columnwidth]{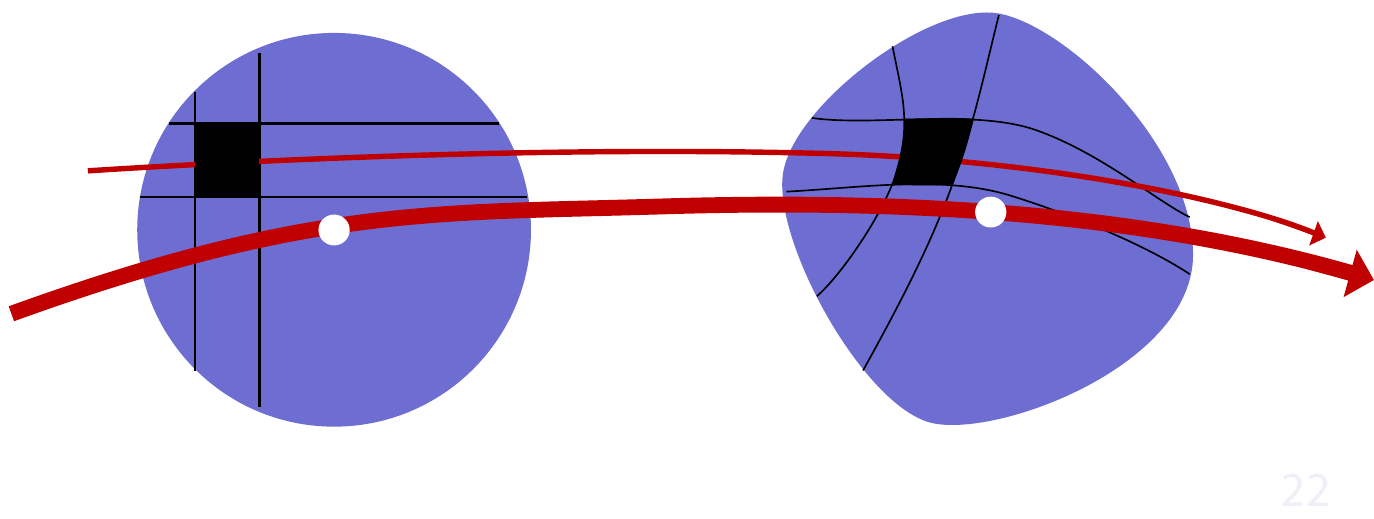}
    \caption{Cartoon of deformable motion, and the  development of a nontrivial metric. As the wavepacket moves, a given shape element both moves relative to the center-of-mass coordinate, and distorts locally due to the presence of the neighboring shape elements. }
    \label{fig:intuitive}
\end{figure}

To be concrete, for the majority of the discussion, we will focus on the dc-conductivity. We note, however, that the same paradigm can also be applied to other response functions. We set the stage by pointing out that perturbation theory probes successively higher derivatives of the quasiparticle motion in terms of the applied perturbation. For example, the current density $\bm{j}$ is related to the applied electric field(s) $\bm{E}$ as follows,
\begin{align}
    j_a&=\sigma^{(1)}_{ab}E_b+\sigma^{(2)}_{abc}E_bE_c+\dots\\
    \sigma^{(n)}_{abc\dots}&\overset{?}{=}
    \frac{1}{n!}
    \frac{D^n j_a}{D E_b D E_c \dots}.
    \label{eq:genderiv}
\end{align}
While formally correct, such a definition of response coefficients has so far not been considered particularly useful, not at least because the nature of the derivative $D$ in Eq.~\eqref{eq:genderiv} is not known.
We find that the Berry connection and the associated transport phenomena have a classical counterpart in the gauge theory of classical deformable bodies, and that the latter theory transparently and straightforwardly implies an emergent curved space in which the body moves. 
This alternative description is visualized in Fig.~\ref{fig:intuitive}: Extended excitations suffer deformations as they move, which leads to a nontrivial effective metric for the center-of-mass coordinate. 
Similarities in the formal definitions, and also the form of the response functions and the closeness of the phenomenology in both descriptions indicate that a canonical response theory in flat space is dual to a gauge theory of deformable bodies.
Since the former description is quantized, while the latter takes place in an effective curved space, we can conclude that the perturbation theory of quantum transport describes quasiparticle motion that takes place in an emergent curved spacetime. 
To repeat, the emergent metric is nontrivial in space\emph{time}, and not only space. Essentially, the deformable wavepacket not only traverses space but also probes remote bands at different energies via virtual transitions. Indeed, we have recently found a mixed axial-gravitational anomaly which supports this conclusion~\cite{Holder2021a}. 

In saying this, at the same time we stress that all the technical details to study these phenomena are fully contained in the standard perturbation theory based on the Kubo formalism in flat space, although  the various components of the response functions have not been recognized before as elements of a motion in curved spacetime. 
Thus, by analyzing these preexisting expressions in the light of the gauge theory of deformable bodies, it will become possible to make predictions about the associated description in a curved manifold.

As the second main result, we find that the anomalous velocity arises because for a deformable body, the elements inside the body can acquire a secular motion compared to the center-of-mass coordinate. This secular motion cannot be captured by global rotations. One can imagine it like picking up a filled bathtub and moving it around erratically; the water inside the bathtub will slosh and tumble and exert additional forces on the carrier.
Most importantly, this picture predicts that genuine new anomalous components of the center-of-mass motion appear at every order in perturbation theory/for each successive time derivative along the quasiparticle trajectory.
For example, for the dc-conductivity we suggest the pattern
\begin{align}
    \sigma^{(1)}&\sim v_{\mathrm{ds}}+v_{\mathrm{bc}}
    \\
    \sigma^{(2)}&\sim a_{\mathrm{ds}}+a_{\mathrm{bc}}+a_{\mathrm{gr}}
    \\
    \sigma^{(3)}&\sim b_{\mathrm{ds}}+b_{\mathrm{bc}}+b_{\mathrm{gr}}+b_{\mathrm{jk}},
\end{align}
where $v_{\mathrm{ds}},a_{\mathrm{ds}},b_{\mathrm{ds}}$ denotes the dispersive velocity contribution or respectively its momentum derivatives, in the same way $v_{\mathrm{bc}}$ denotes the anomalous velocity associated with the Berry curvature, $a_{\mathrm{gr}}$ is the anomalous acceleration induced by the metric and $b_{\mathrm{jk}}$ the anomalous jerk component.
We note that the first two lines conform with known expressions~\cite{Parker2019,Holder2020}, whereas the third line is our prediction.

Another way of stating the same observation goes as follows. Due to the internal degrees of freedom afforded by deformations of the wavepacket, in each successive order in perturbation theory, a new band structure object arises which characterizes certain aspects of the excited state. In sum, the infinite series of interband quantities by construction contains the complete information about the adiabatic response of the system. Therefore, the interband quantities, together with the information about the ground state, completely characterize the band structure. This presents a tractable generalization of the statement that in gapless systems the ground state is completely determined by the Fermi surface.

\section{Theory of deformable bodies}
\begin{figure}
    \centering
    \includegraphics[width=.8\columnwidth]{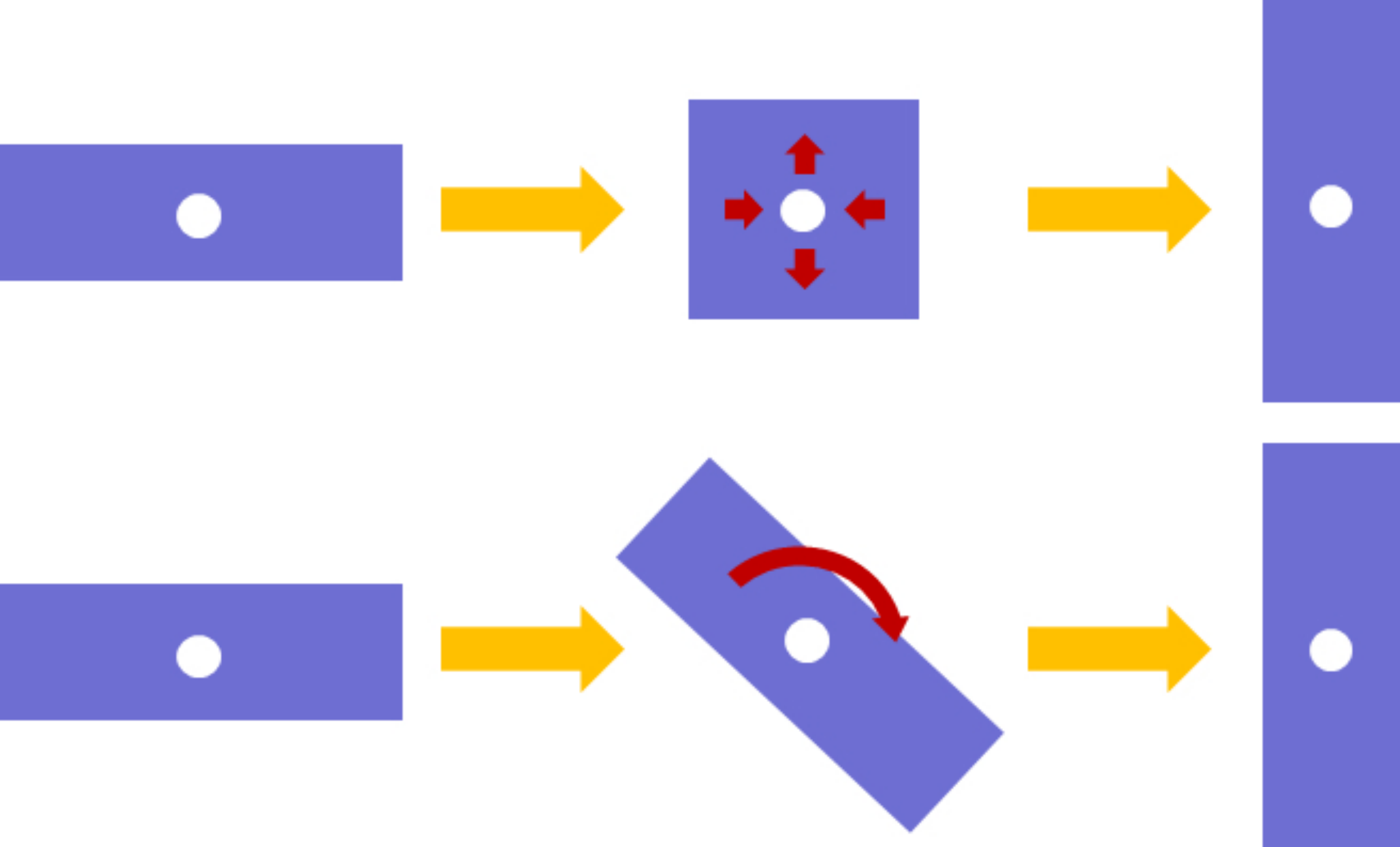}
    \caption{Necessity of a gauge potential. Deformable bodies can mimick rotations by a series of rotationless deformations. A given state is thus not uniquely connected to the preceding one, and uniqueness is only restored after the introduction of a gauge potential.}
    \label{fig:fig2}
\end{figure}
The motion of deformable objects is well-studied~\cite{Littlejohn1997}. Let's define a classical, extended object by a number $N$ of mass elements $m_\alpha(\bm{R}_\alpha)$ at real space positions $\bm{R}_\alpha$. In the following discussion for simplicity we will use a discrete index $\alpha$, with the generalization to continuous variables being straightforward. The object's motion can be analyzed in three steps. Firstly, there is a translation of the center-of-mass coordinate $\bm{R}_{cm}$, which can be subtracted out. The resulting coordinate system is known as the \emph{configuration space}, which is sensitive to both changes of the object's shape and changes of orientation, but not translations.
The second step is the subtraction of rotations.
However, it is a fundamental tenet of the motion of deformable bodies that shape transformations and rotations are not completely separable. This is because an object can change shape in a sequence of steps that do not involve any rotation (i.e. the angular momentum is zero) and end up in a configuration which has the same shape, but is rotated (cf. Fig.~\ref{fig:fig2}).
It is therefore necessary to specify a reference orientation for each shape, thereby introducing a gauge convention.
The \emph{shape space} of an object can then be described by a collection of shape coordinates $q^\mu$ which are invariant under proper rotations. 
In total, the $3N$ coordinates $\bm{R}_\alpha$ are transformed into $\bm{R}_{cm}$, three Euler angles $\theta_\alpha$ rotations and $3N-6$ shape coordinates $q^\mu$.
We note that the rotations contain both global (trivial) rotations of the kind known from rigid bodies, and gauge dependent rotations which are determined by the choice of reference orientation for a given shape.
Therefore, a slightly more transparent system of coordinates than configuration space are the orientational, mass-weighted Jacobi coordinates $\bm{\rho}$, which are defined excluding proper (global) rotations via,
$(\mathsf{R}\bm{\rho})_\alpha=\sqrt{\mu_\alpha}\sum_{\beta=1}^{n}\mathsf{T}_{\alpha\beta}\bm{r}_{\beta}$,
where $\bm{r}_\alpha=(\bm{R}_\alpha-\bm{R}_{cm})$ are the relative coordinates, and $\mathsf{R}$ is a global rotation. Furthermore, $\mu_\alpha$ is a reduced mass and  $\mathsf{T}_{\alpha\beta}$ implements the subtraction of the translational part of the kinetic energy~\cite{Littlejohn1997}. This construction might seem unnecessarily complicated, if it weren't for the great simplification that results for the dynamical quantities and the equation of motion. 
The gauge potential mentioned earlier translates between the mass weighted orientational coordinates $\bm{\rho}_\alpha$ and the shape coordinates $q^\mu$ and is defined by
\begin{align}
\bm{A}_\mu=\mathsf{M}^{-1}\sum_{\alpha}^{N-1}\bm{\rho}_\alpha\times\frac{\partial\bm{\rho}_\alpha}{\partial q^\mu},
\end{align}
where the inertia tensor is $\mathsf{M}_{ij}(q)=\sum_{\alpha}^{N-1}|\bm{\rho}_\alpha|^2\delta_{ij}-\rho_{\alpha i}\rho_{\beta j}\equiv
\sum_{\alpha}^{N}m_\alpha(|\bm{r}_\alpha|^2\delta_{ij}-(\mathsf{R}^{-1}\bm{r})_{\alpha i}(\mathsf{R}^{-1}\bm{r})_{\beta j})$.
One further introduces a metric
\begin{align}\label{eq:classicalmetric}    g_{\mu\nu}=\sum_{\alpha}^{N-1}
\frac{\partial\bm{\rho}_\alpha}{\partial q^\mu}
\frac{\partial\bm{\rho}_\alpha}{\partial q^\nu}
-\bm{A}_\mu \mathsf{M} \bm{A}_\nu.
\end{align}
For a potential $V$, the classical Langrangian which describes the internal motion of the deformable body can then be expressed in terms of $q^\mu$ as
\begin{align}
    \mathcal{L}
    &=
    \tfrac{1}{2}(\bm{\omega}+\bm{A}_\mu \dot{q}^\mu)
    \mathsf{M}
    (\bm{\omega}+\bm{A}_\mu \dot{q}^\mu)
    \notag\\&\quad
    +\tfrac{1}{2}g_{\mu\nu}\dot{q}^\mu\dot{q}^\nu
    -V(q)
    \label{eq:classicalL}
\end{align}
from which we read off the angular momentum 
$\bm{L}=\cdot{M}(\bm{\omega}+\bm{A}_\mu \dot{q}^\mu)$. Note that we use Einstein summation convention for repeated shape indices (denoted with greek letter $\mu,\nu,\dots$) and spatial indices (lowercase latin letters).
For $V=0$ and $\bm{L}=0$, the equation of motion for the Lagrangian Eq.~\eqref{eq:classicalL} is a that of geodesic,
\begin{align}
  \frac{D \dot q^\mu}{Dt}&=\ddot q^\mu+\Gamma^\mu_{\;\;\sigma\tau}\dot q^\sigma \dot q^\tau=0,
\end{align}
with the usual definition of the Christoffel symbol $\Gamma^\mu_{\;\;\sigma\tau}=\tfrac{1}{2}g^{\mu\nu}(\frac{\partial g_{\nu\sigma}}{\partial q^{\tau}}+
\frac{\partial g_{\nu\tau}}{\partial q^{\sigma}}-
\frac{\partial g_{\sigma\tau}}{\partial q^{\nu}})$.
It describes the motion of the shape elements in the non-Euclidian space which emerges due to deformations of the body.
Furthermore, one defines the Coriolis vector $\bm{B}_{\mu\nu}=\partial_\mu \bm{A}_\nu-\partial_\nu \bm{A}_\mu-\bm{A}_\mu\times \bm{A}_\nu$, which parametrizes the additional forces which arise due to the secular motion of the shape elements.
We remark that both for $\bm{A}_\mu$ and $\bm{B}_{\mu\nu}$, it is sometimes beneficial to use the corresponding antisymmetric tensor, with notation $\mathsf{A}_\mu$ and $\mathsf{B}_{\mu\nu}$.

In summary, by introducing shape coordinates, the motion is subdivided into a nontrivial connection in configuration space, called $\bm{A}$, and a nontrivial metric in shape space, encoded in the Christoffel symbol $\Gamma$.
Therefore, the covariant time derivative of a general function with with spatial indices and shape indices acquires additional terms of the type $-\bm{A}_\mu\dot{q}^\mu$ for each spatial index and terms of the form $\Gamma^\mu_{\sigma\tau}\dot{q}^\sigma$ for each shape index. 
This constitutes the mathematical origin of the anomalous components of the center-of-mass motion.

\section{Response functions and generalized derivatives}
In the beginning we introduced the notion that perturbation theory probes successively higher time derivatives of the quasiparticle motion. Utilizing the gauge formulation of classical deformable bodies, we now make this statement precise.
A subtle point is the fact that the quasiparticle dynamics depends on the Berry connection $\bm{\mathcal{A}}$, a complex gauge field which is not identical to the real-valued classical gauge potential $\bm{A}$. We will disregard this complication at first and come back to it in the end.

Consider a simple band Hamiltonian in first quantized notation
\begin{align}
    H&=-\sum_i \tfrac{\hbar^2 \nabla_i^2}{2 m_i}+V(r)
    \label{eq:primH}
\end{align}
where $V(r)$ is a periodic potential and the coordinates are placed such that they fill the unit cell of the lattice. 
The eigenstates of the system at equilibrium are given by a number of wavefunctions $\psi_{n\bm{k}}(r)$ with band index $n$ and momentum $\bm{k}$, which are periodic between unit cells up to a Bloch phase.
We will mostly suppress the dependence on $\bm{k}$ and denote the cell periodic part of the eigenfunctions in momentum space as $|m\rangle$, which means that the Berry connection is $\mathcal{A}_{a,mn}=\langle m|i\partial_{k_a} |n \rangle$.

Enforcing periodicity between unit cells can be viewed as a labeling of $\psi_n(r)$ in a coordinate system which disregards both translations and rotations, i.e. it represents a choice of certain shape coordinates. 
This makes sense because in equilibrium, the wavefunctions carry neither a global center-of-mass motion nor do they perform a global rotation.
Therefore, all considerations within perturbation theory proceed under the assumption that the initial and final state have zero velocity and zero angular momentum.
We emphasize that by saying that the angular momentum is zero, we do not at all mean that orbitals with finite angular momentum are absent or that there are no spin degrees of freedom. The statement is that in the ground state there is neither a macroscopic current flowing nor are there microscopic current loops, which is a standard requirement for the ground state~\cite{Nenciu1991}. 
Let us first assume that the choice of labeling of a unit cell is preserved continuously as a function of time, which might be seen as the selection of a \emph{consistent frame}.
Then, the time evolution under a perturbation will result in the displacement and distortion of the chosen shape ('unit cell') as a function of time.
\begin{figure}
    \centering
    \includegraphics[width=\columnwidth]{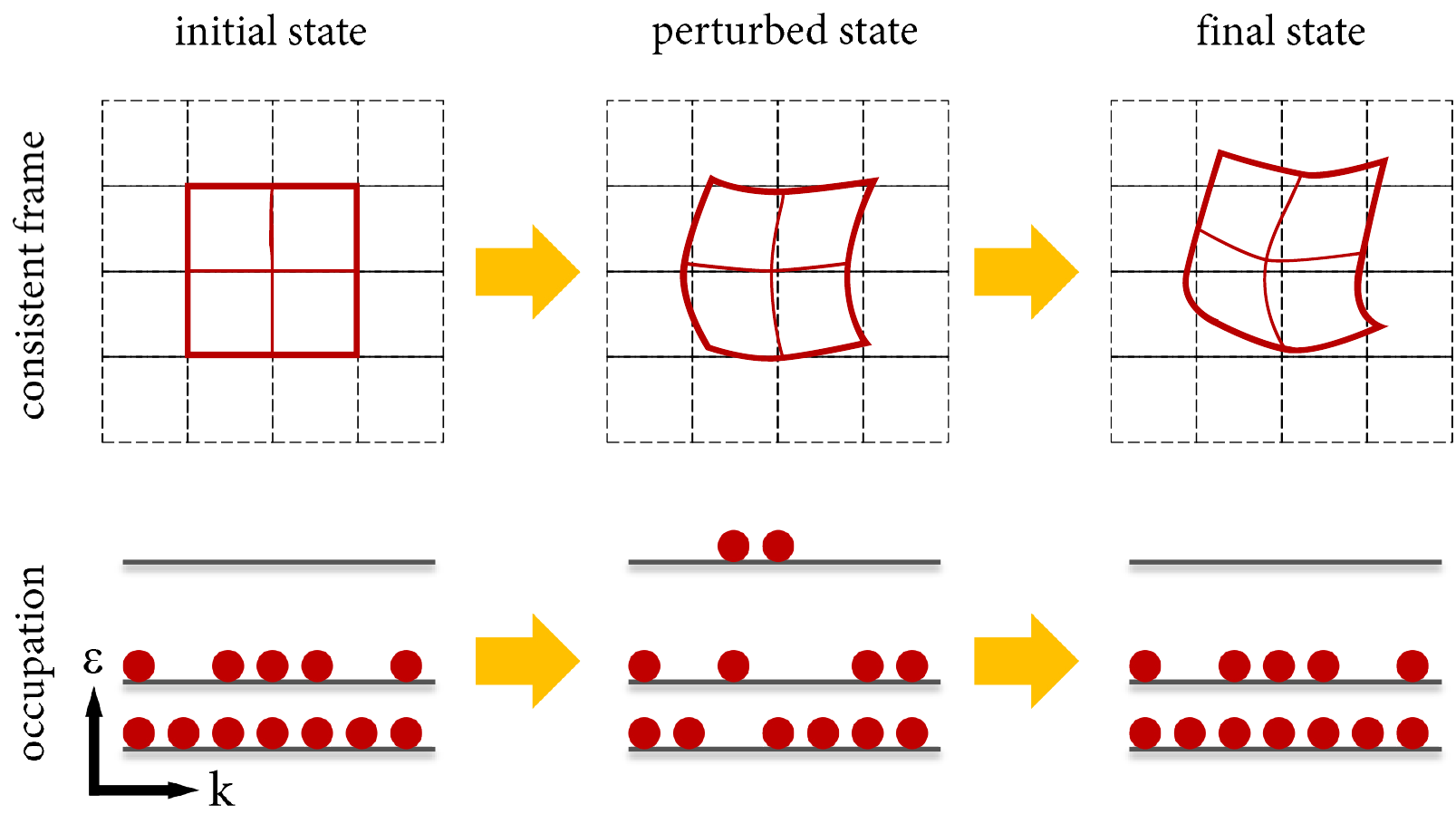}
    \caption{ (a) Implementation of shape coordinates in a periodic system. The straightforward implementation where each shape element is transported continuously between time steps (continuous frame)  will fail to restore the inital state manifestly. It is therefore beneficial to use a fixed frame, where shape elements are redefined after each time step. (b) In perturbation theory, initial and final state are by construction identical, but intermediate states can involve real or virtual excitations.}
    \label{fig:fig3}
\end{figure}
At this point, it is important to recall that adiabatic perturbation theory considers any incurred motion as a closed cycle: The initial state is perturbed by introducing some external force, which is then removed again, which returns the system to its initial state.
Therefore, at the end of the cycle, the wavefunction has to resemble the equilibrium wavefunction (up to phases). However, the final shape in the consistent frame may differ from the starting shape (cf. Fig.~\ref{fig:fig3}). Obviously, all accumulated differences are an artefact of the choice of reference frame and can be undone by a relabeling of the volume elements. 
This shows that under a perturbation the consistent frame is counter-intuitive in that it does convey a notion of translation and rotation. In a periodic system, a suitable choice of coordinates is therefore one that at each point in time reassigns coordinates such that $\psi_n(r)$ only resides within a fixed unit cell. Of course, this can immediately be identified as the standard choice for the treatment of periodic systems.
In the chosen \emph{fixed frame}, all motion is incurred by gauge transformations. 
Therefore, the usage of a fixed unit cell in perturbation theory is equivalent to a frame of reference that amounts to shape coordinates.
As formally it does not make a difference whether gauging is a passive or an active transformation, we conclude that response theory for periodic systems implements the nature of shape transformations in a fixed frame.
To be clear, we reiterate that a choice of shape coordinates together with a gauge convention also fixes the configuration space, which implements the spectral flow in a periodic system in which the quasiparticle response manifests itself, but implicitly so.

How does this procedure look like in practice? 
For notational convenience, we retain the discrete label $q_\mu$ for the continuous spatial variables. 
Next, we absorb the band index $n$ into $\mu$. This hides the multiband nature of the wavefunction, i.e. the fact that the gauge field may be non-diagonal in the band index $n$. 
To recover the band notation, one can reinsert for $\bm{A}$ a nontrivial irreducible representation in band space, but this step does not affect any of the formal developments which we elucidate now.

In the following we will discuss the motion in momentum space, with the chosen shape being the Brioullin zone. The shape coordinate in momentum space is called $k^\mu$.
The current density $\bm{j}$ is a function of the perturbing electric field $\bm{E}$, subject to the equilibrium condition $\bm{j}(\bm{E}=0)=0$.
Next, we construct the covariant derivative with respect to shape transformations as a function of an applied electric field. We reiterate that the regular derivative is not very useful by itself because it encodes the change from $\bm{j}(\bm{E}_0)$ to $\bm{j}(\bm{E}_0+\delta \bm{E})$, both of which correspond to dissimilar body frames at points $k^\mu$ and $k^\mu+\delta k^\mu$. In obvious generalization of the time dependence of deformable objects which we discussed beforehand, it is
\begin{align}
    &\frac{D j_a}{D E_b}\biggr|_{\bm{E}=0}=
    \frac{\partial j_a}{\partial E_b}\biggr|_{\bm{E}=0}
    \notag\\
    &\quad
    -\Bigl(\mathsf{A}_{\mu,ad}j_d\frac{\partial k^\mu}{\partial E_b}
    +\mathsf{A}_{\mu,bd}j_a\frac{\partial k^\mu}{\partial E_d}
    \Bigr)
    \biggr|_{\bm{E}=0}
    \label{eq:lincurrent}
\end{align}
We emphasize that expression like Eq.~\eqref{eq:lincurrent} are by themselves of little use unless the field dependence of the polarization is explicitly constructed by the means of a microscopic model. Nevertheless, because we already know the form of the linear-response conductivity from perturbation theory~\cite{Parker2019,Holder2020}, we propose the following relation between the classical current density and the shape coordinates
\begin{align}
    \mathsf{\tilde A}_{\mu,mb}&=V^{-1}_{BZ}\frac{\hbar}{e^2}j_a\frac{\partial k^\mu}{\partial E_b}.
    \label{eq:shapecurrent}
\end{align}
where $V_{BZ}$ is the volume of the Brillouin zone and the tilde cautions us that the linear conductivity in Eq.~\eqref{eq:lincurrent} only makes use of the skew-symmetric entries in the gauge potential $A$.
We point out that the sum over shape index $\mu$ in Eq.~\eqref{eq:lincurrent} expands into a sum over non-equal shape indices after restoring the band index, as mentioned before. This is because a momentum can only be defined up to a reciprocal lattice vector due to the periodicity of the lattice, meaning that the commutator in Eq.~\eqref{eq:lincurrent} acquires a matrix structure in shape indices. 

The identification Eq.~\eqref{eq:shapecurrent} is reminiscent of the modern definition of the polarization density $\bm{P}$ in terms of the Berry connection $\bm{\mathcal{A}}$~\cite{Resta2007}
{\allowdisplaybreaks
\begin{align}
\bm{P}&=-e\int_{\bm{k}} \Re \bm{\mathcal{A}}\\
\partial_t\bm{P}
&=-\tfrac{e^2}{\hbar}\int_{\bm{k}} (\bm{E}\cdot\partial_{\bm{k}})\Re \bm{\mathcal{A}}
\\
&\rightarrow
\tfrac{e^2}{\hbar}\int_{\bm{k}} (\partial_{\bm{k}}\cdot\bm{E})\Re \bm{\mathcal{A}},
\label{eq:quantumcurrent}
\end{align}}
where the equation of motion for crystal momentum $\hbar \bm{\dot k}=e\bm{E}$ was used to replace $\partial_t\Leftrightarrow\tfrac{e}{\hbar}\bm{E}\cdot\partial_{\bm{k}}$. 
We note that the last line which includes a derivative in terms of the (external) electric field does not have a clear meaning in the usual perturbation theory, because $\bm{E}$ is considered an independent variable. 
If there were a relation local in momentum space that reads $\Re \bm{\mathcal{A}}\sim V_{BZ}^{-1}\tfrac{\hbar}{e^2}(\partial_{\bm{E}}\cdot\bm{k})\bm{j}$, then Eq.~\eqref{eq:quantumcurrent} would immediately follow as a corollary. However, such a relation has not been found. 
In comparing the definition of the current in the framework of classical deformations, Eq.~\eqref{eq:shapecurrent}, we notice some additional differences: 
According to Eq.~\eqref{eq:shapecurrent}, the derivative $\partial k^\mu/\partial E_a$ contains additional structure due to the appearance of a shape index. This corresponds to the statement that $\bm{\dot k}$ is not necessarily collinear to $\bm{E}$, which is exactly what is captured by a covariant derivative but not an instantaneous derivative. 
We also find that the gauge potential pseudovector $\bm{A}$ is orthogonal to both the current and the direction of displacement of each shape element. Again, a statement like this is not known for the Berry connection $\bm{\mathcal{A}}$.
Therefore, $\bm{A}$ and $\bm{\mathcal{A}}$ are not simply related. We attribute these differences to the more general starting point which can be afforded by the gauge theory of deformable bodies: In Eq.~\eqref{eq:lincurrent} for the linear conductivity, all quantities are defined incorporating the complete response of the system, and they are only evaluated at zero perturbation. In contrast, in perturbation theory all quantities appearing on the right hand side are ground state properties of the system.

Nonetheless, the theory of deformable bodies offers a couple of exciting new insights. 
First, we give a new interpretation of the non-Abelian properties of the gauge potential. To this end, we recall that Coriolis tensor $\mathsf{B}_{\mu\nu}$ is nonzero only if the body performs some kind of rotation. Obviously, in the perturbation theory no rotation is taking place, thus $\mathsf{B}_{\mu\nu}=0$. However, this only enforces that $\mathsf{B}_{\mu\nu}=\partial_\mu \mathsf{A}_\nu-\partial_\nu \mathsf{A}_\mu-[\mathsf{A}_\mu,\mathsf{A}_\nu]=0$, whereas $[\mathsf{A}_\mu,\mathsf{A}_\nu]$ by itself can be nonzero. Indeed, in the canonical formalism using Bloch states, the response functions receive corrections from the Berry curvature $\Omega_{ab}=\partial_a \mathcal{A}_b-\partial_b \mathcal{A}_a$, while it holds that $\partial_a \mathcal{A}_b-\partial_b \mathcal{A}_a=i[\mathcal{A}_a,\mathcal{A}_b]$~\cite{Holder2020}. Thus, while no global rotation takes place, the  wavefunction is deformed in such a way during the adiabatic perturbation that it takes the appearance of an \emph{internal} rotation. 
This conclusion is actually unavoidable when considering that a wavefunction which carries the (discrete) spatial symmetries of the periodic lattice cannot possibly perform a continuous rotation unless it is not a rigid body, i.e. the rotation is actually a deformation.
The explanation that $\mathsf{B}_{\mu\nu}=0$ also fully rationalizes why the quasiparticle motion couples to the Berry curvature $\Omega_{ab}=\partial_a \mathcal{A}_b-\partial_b \mathcal{A}_a$ and not to the full Coriolis tensor: The 'rotation' of the wavefunction is actually a deformation. 
In the literature this subtle point has previously been treated by creating a distinction between Abelian and non-Abelian Berry curvature when talking about diagonal and off-diagonal matrix elements in terms of band indices~\cite{Xiao2010}, which is problematic because the Berry curvature is always non-Abelian, and the same definition should apply to all matrix elements.
Instead, we argue that the absence of any Coriolis force in the laboratory frame requires the presence of the Berry curvature term in the response functions, in order to compensate for the deformation of the wavepacket.

Secondly, we point out that the quantum metric , defined as~\cite{Marzari1997,  Resta2011,Marzari2012}
\begin{align}\label{eq:quantummetric}
    \mathcal{G}_{ab}
    &=\sum_{n \in occ.}
    \Re[\langle \partial_{k_a} n |\partial_{k_b} n\rangle
    -\sum_m\mathcal{A}_{a,nm}\mathcal{A}_{b,mn}]
\end{align}
is conceptually very similar to the metric in shape space [cf. Eq.~\eqref{eq:classicalmetric}].
Namely, both contain the product of two derivatives of 'half-densities' over shape space, and in both the vertical part of the motion along the fiber bundle~\cite{Littlejohn1997} is subtracted by the analogous term in each formalism. This is equivalent to saying that the metric is dynamically generated by the moving shape elements, minus any components that originate from finite angular momentum.
However, we emphasize that Eq.~\eqref{eq:classicalmetric} is a scalar in real space, while Eq.~\eqref{eq:quantummetric} defines a tensor.
Classically, the metric encodes the effective mass of the shape elements, which in perturbation theory we interpret as the local effective geometry that is enforced by the neighboring states.

It is important to keep in mind that the appearance of deformations is concomitant with the appearance of anomalous terms in the semiclasscial motion, which in turn is equivalent to saying that the quasiparticle responses involves (virtual) interband processes (cf. Fig.~\ref{fig:fig3}).
This is certainly true for the anomalous velocity, which originates from the Berry curvature, an object which explicitly contains interband terms in its definition. More recently, similar observations were put forward for the anomalous acceleration~\cite{Holder2020}.

In summary, we propose Eq.~\eqref{eq:lincurrent} as the definition of the linear conductivity when using shape coordinates, which coincides with the classical current due to perturbation theory if we additionally impose Eq.~\eqref{eq:shapecurrent}. Notably, it allows for an intuitive understanding of the various pieces as properties of the quasiparticle response, rather than stating the response in terms of an ever increasing number of matrix elements.
A construction of higher order response coefficients is possible through repeated application of the affine connection. 
For example, at second order in the electric field, the same prescription yields 
\begin{align}
    &\frac{D^2 J_a}{D E_b D E_c}=
    \frac{\partial^2 J_a}{\partial E_b \partial E_c}
    \notag\\&\quad
    +V_{BZ}\frac{e^2}{\hbar}
    \biggl[
    -\frac{\partial}{\partial E_a}\tilde\Omega_{bc}
    -\frac{\partial}{\partial E_b}\tilde\Omega_{ab}
    \notag\\&\quad
    +\mathsf{A}_{\mu,ad}\frac{\partial k^\mu}{\partial E_d}\tilde\Omega_{bc}
    +\mathsf{A}_{\mu,bd}\frac{\partial k^\mu}{\partial E_a}\tilde\Omega_{dc}
    +\mathsf{A}_{\mu,cd}\frac{\partial k^\mu}{\partial E_a}\tilde\Omega_{bd}
    \biggr]
    \label{eq:seccurrent}
\end{align}
where as before, all quantities are to be evaluated at ${\bm{E}=0}$ and we introduced the shorthand $\tilde\Omega_{ab}=[\mathsf{A}_\mu, \mathsf{\tilde A}^\mu]_{ab}$, which is a scalar in shape space and a tensor in spatial coordinates.
We thus see that at every order in the response coefficients, an additional anomalous term is introduced by the covariant derivative. The benefit in the language of Eq.~\eqref{eq:seccurrent} is that at each order the response is expressed in terms of the previously created object. For example, at second order, there are direct derivatives of the current, direct derivatives of $\Omega$ and additionally the commutators of $\Omega$. 
These insights give credence to our semiclassical interpretation of response coefficients as successively higher derivatives of the response function with respect to the applied perturbation. Essentially, the n-th order response quantifies the n-th order monomial in a polynomial that approximates the quasiparticle trajectory of an extended and thus deforming object.
At the same time, the formulation of Eq.~\eqref{eq:seccurrent} is impractical because it requires at each order the construction of a new observable, which seems exceedingly difficult without prior knowledge. 

One might be concerned that the discussion so far is lacking concrete applicability because both formalisms are so different, and also because we did not establish a mathematical prescription to translate between them.
Therefore, we now showcase with the example of the second order conductivity how the language of wavepacket deformation can be utilized to give intuitive understanding to various expressions in perturbation theory, thereby guiding us to new predictions about concrete material properties.

\section{Second order conductivity and anomalous acceleration}
The diagrammatic approach to second order optical response has been developed recently~\cite{Parker2019,Holder2020}. We briefly state the main results here~\cite{Kaplan2021a}
\begin{align}
	\sigma_{ab;c}^{(2)}&=
	\frac{2e^3}{\hbar^2}
	\int_{\bm{k}} \sum_{n\in occ.} \tau^2 (\partial_{k_a} \partial_{k_b} \partial_{k_c} \varepsilon_n)
	\notag\\&\quad
	+\tau (\partial_{k_a} \Omega_{bc,n} + \partial_{k_b} \Omega_{ac,n})
	+\partial_{k_c} G_{ab,n},
	\label{eq:secorder}
\end{align}
where $n$ is the band index, $\hbar\varepsilon_n(\bm{k})$ is the dispersion, $\Omega_{ab,n}(\bm{k})=\sum_{m\neq n}i(\mathcal{A}_{a,nm}\mathcal{A}_{b,mn}-\mathcal{A}_{b,nm}\mathcal{A}_{a,mn})$ is the Berry curvature tensor, and $G_{ab,n}(\bm{k})=\sum_{m\neq n}(\mathcal{A}_{a,nm}\mathcal{A}_{b,mn}+\mathcal{A}_{b,nm}\mathcal{A}_{a,mn})/(\varepsilon_{m}-\varepsilon_n)$ has units of area per energy, which can be interpreted as a density of states for deformations.
$\tau$ denotes the quasiparticle relaxation rate.
Given the discussion of the previous section, we identify the second order conductivity as a probe of the acceleration of a wavepacket under an applied electric field. 
This means in particular that we attribute to each of the three contributions in Eq.~\eqref{eq:secorder} a distinct physical process that accelerates the wavepacket, i.e. they encode different ways in which the velocity vector of the center-of-mass coordinate changes.
The first term in Eq.~\eqref{eq:secorder} is the dispersive acceleration, which acts on the dispersive velocity vector~\cite{Watanabe2020}.
Secondly, there is the acceleration of the self-rotation, i.e.\ the change of the Berry curvature in the applied field. In perturbation theory it takes the form of a momentum derivative of the Berry curvature, therefore this term is also known as the Berry curvature dipole~\cite{Sodemann2015,Morimoto2016}.
As expected from the formalism for deformable bodies, the third term in Eq.~\eqref{eq:secorder} contains a genuine new term (i.e.\ it is not a momentum derivative of any of the previous terms)~\cite{Gao2014,Gao2019}.
According to Eq.~\eqref{eq:seccurrent}, it represents the secular motion of the anomalous velocity, quantifying to which amount the Berry curvature rotates in the applied field.
This can be thought of as a type of Magnus effect, whereby a rotating object is itself pushed to the side by the flow (or its dipole vector is rotated if it is a vortex-antivortex pair).
Indeed, a similar but nonidentical Magnus Hall effect appears when a potential gradient is applied perpendicularly to the normal bias~\cite{Papaj2019}. 

The role of deformations becomes more pronounced when considering the bulk photovoltaic effect, which is a resonant dc-current created by an electric field of finite frequency. This effect appears in two qualitatively distinct forms, known as the injection current and shift current~\cite{Sipe2000,Morimoto2016a,Cook2017,Parker2019,Holder2020,deJuan2020}. 
On one hand, the expressions are now explicitly dependent on a finite frequency, making it more difficult to find classical counterparts for the various terms. 
On the other hand, every deformation leads to a novel type of anomalous acceleration, which can be interpreted quite straightforwardly.
\begin{figure}
    \centering
    \includegraphics[width=.55\columnwidth]{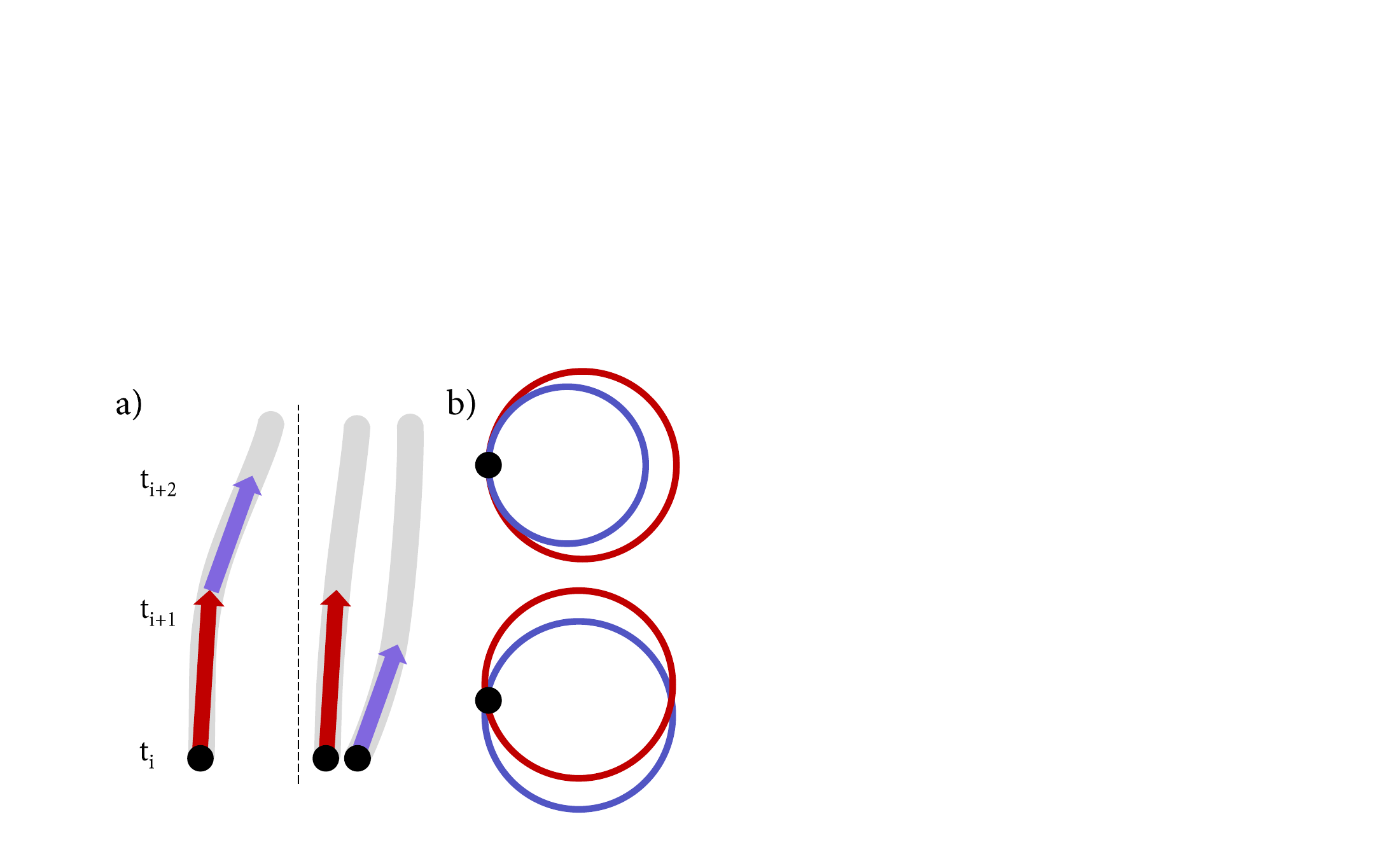}
    \caption{Visualization of the two leading causes of an anomalous acceleration in resonant second order optical response. (a) To obtain an acceleration, the quasiparticle trajectory is evaluated at three points. Equivalently, one can examine the differences between two closeby trajectories. (b) Based on the two velocity vectors, one can draw two different types of closed trajectories, one that differ in curvature (top), and one that is not locally parallel (bottom).}
    \label{fig:fig4}
\end{figure}
To this end, let us imagine an acceleration as two vectors (Fig.~\ref{fig:fig4}). By placing both vectors in close proximity, we create a situation which closely resembles the parallel transport problem along a manifold.
In accordance with perturbation theory, we next draw closed trajectories (Fig.~\ref{fig:fig4}) for both vectors. If both circles differ in length, we attribute this to the metric. If they differ in direction, the phenomenon is quantified by Christoffel symbols. 
As a matter of fact, in a low frequency expansion, this kind of mapping has been suggested recently for injection and shift current~\cite{Ahn2020}. Namely, the injection current has been associated with the quantum geometric tensor, and the shift current with (symplectic) Christoffel symbols which originate from this geometric tensor.

Several predictions can be made from the association of the second-order conductivity with the anomalous acceleration of a deformable body.
Most importantly, the phenomenon of the shift current was originally attributed to a real-space displacement of the wavefunction~\cite{vonBaltz1981}, which cannot rationalize that the shift current can still be very large in systems with a high quasiparticle mass~\cite{Kaplan2021}. In contrast, an explanation involving the anomalous acceleration instead requires a strong effect of virtual transitions from the nearby bands. Indeed, the shift current is very large in Weyl semimetals~\cite{Cook2017} and in materials with flatband dispersion~\cite{Kaplan2021}, both of which prominently feature interband transitions.

It is also possible to explain much easier when and why the injection current appears, given that it is essentially a case of what could be termed 'dynamical antilocalization'~\cite{Holder2020}: It appears in situations where time-reversal symmetry is broken, either by the system or the incident light.
Indeed, while current injection has historically been considered as a response to circular polarized light, in magnetic materials the injection current appears also for linear polarized light~\cite{Zhang2019}. 
Furthermore, according to Fig.~\ref{fig:fig4}, injection and shift have a closely related physical origin. Case in point, for energies just below the band gap in a semiconductor or insulator, both shift and injection current cancel each other, unless there is a mismatch of relaxation rates in the system~\cite{Kaplan2020}.

Finally, we recently succeeded in showing that the third term in Eq.~\eqref{eq:secorder} is the result of a mixed axial-gravitational anomaly~\cite{Holder2021a}.
Thanks to our understanding of anomalous terms as deformations, we can immediately identify as the origin of the anomaly a local violation of charge conservation due to virtual excitations, which is equivalent to a change of the local geometry of the Fermi surface. The latter must entail the emergence of a nontrivial metric.
Because the mixed axial-gravitational anomaly requires both the spatial and the temporal components of the Riemann tensor to be nontrivial, the presence of this anomaly gives credence to our claim that electrons move in an emergent curved spacetime, not only in curved space.

Other examples of quasiparticle response have been discussed in the literature which we believe can be subsumed into the same phenomenology of deforming wavepackets.
On one hand, this encompasses n-th order responses. Here we mention in particular the recent observation that the third order electrical conductivity~\cite{Fregoso2019,Kumar2021} - a four index object - can be expressed with the help of a Riemann curvature tensor~\cite{Ahn2021}.
On the other hand, many transport problems have been investigated with respect to smooth spatial variations, i.e. the corresponding response coefficient is analyzed at nonzero momentum. 
For example, in the case of the nonreciprocal directional dichroism, it was found that it depends on the quantum metric dipole~\cite{Gao2019a}. 
Similarly, the leading order correction in the momentum dependence of the linear Hall conductivity carries the same symplectic Christoffel symbols that were also reported for the shift current~\cite{Ahn2020,Kozii2021}. 

In summary, the intuitive picture of higher order responses as (anomalous) higher order derivatives of the center-of-mass coordinate offers a unifying principle to understand, and predict interesting features in response coefficients in terms of the underlying band structure properties.

\section{Some conjectures}

Given the potential far-reaching nature of the presented physical picture, we feel it is appropriate to present two slightly more speculative aspects.
Firstly, we suspect that
all transport coefficients or response functions carry in some form or the other information about the deformation degrees of freedom. 
We have discussed how such can be identified in the second order conductivity, but a similar approach seems feasible in many other instances.
We also point out that we completely sidelined a discussion of deformation effects in terms of an angular momentum expansion. Such is probably adequate and a promising approach. We suspect that it will lead to a Fermi-liquid type of phenomenology for nonlinear transport effects. This hints at the close relationship of the deformation effects for transport discussed here, and suggests that the same mechanism is present when considering a dynamically changing Fermi surface.

This brings us to our second point. We conjecture that metallic quantum criticality~\cite{Belitz2005,Loehneysen2007}, maps to a theory with a dynamically generated gravitational field.
The reason for this claim is that close to criticality, the Fermi surface becomes soft and is easily deformed, with the consequence that the quasiparticle similarly becomes squishy and can dynamically suffer deformations~\cite{Varma2002,Metzner2003,Senthil2008}.
Indeed, like in quantized Einstein gravity~\cite{Goroff1986}, there are concrete indications documented in the literature where metallic quantum critical points become non-renormalizable at second order in perturbation theory~\cite{Abanov2004,Rech2006} or at higher order~\cite{Holder2015}.
Conversely, all known cases of renormalizable quantum critical theories seem to feature a locally linear dispersion where deformations do not impose a tensor structure~\cite{Schlief2017,Lee2018}.
We caution that we do not necessarily see these considerations connected to holographic methods which have also been studied for these systems (see e.g.\ Ref.~\cite{Hartnoll2018} for a review).

\section{Conclusions}
In this work we suggested to view semiclassical quasiparticle motion as the motion of a deformable body, the main implication being that the quasiparticles move in an emergent curved spacetime.
For pedagogical reasons, we demonstrated the similarity between both approaches using a classical description.
By no means we are suggesting that the use of shape coordinates is superior to the use of canonical perturbation theory. 
Rather, the gauge theory of deformable bodies offers a straightforward interpretation to some otherwise puzzling findings. 
Namely, we gave a new reasoning why the Berry curvature appears, and how it introduces an anomalous velocity. We also gave a new interpretation to the quantum metric as the local geometry that is spanned by the shape elements. Furthermore, we argued that genuine new anomalous terms appear in the semiclassical equations of motion at every order in perturbation theory.
Using the second order electrical conductivity, we demonstrated in detail how these concepts can be applied for a concrete response coefficient.
Among the most striking conclusions, we elucidated that our recent discovery of a mixed axial-gravitional anomaly in the anomalous acceleration establishes unambiguously that electrons effectively move in curved spacetime, and not only in curved space, as classical deformable bodies would do.
This result might offer a route to table-top experiments on synthetic gravitational fields.

We were not able to extensively cover how to utilize the quantized formulation of the gauge theory of deformable bodies for response coefficients, which requires the half-densities to be replaced by complex-valued wavefunctions. We believe it is worthwhile to investigate these aspects in detail, because given a formulation in shape coordinates, one could then  explicitly compare the canonical formalism in flat space with the expressions written for a curved manifold.

Finally, we offered a speculative outlook about possible further ramifications of the physical picture presented here, in particular highlighting the applicability of this approach for all types of response functions, and the possible connection to advanced questions in quantum field theory which arise in the description of quantum critical metals. We suggest to investigate this line of reasoning further because it might constitute a road towards realizing quantum gravity analogues in condensed matter systems.

\begin{acknowledgements}
We thank 
Tabea Heckenthaler,
Johannes Hofmann, 
Roni Ilan,
Daniel Kaplan,
Raquel Queiroz and
Binghai Yan
for enlightening discussions.
\end{acknowledgements}

\appendix

%

\end{document}